\documentclass[reprint,superscriptaddress,amsmath,amssymb,aps,prl,floatfix]{revtex4-1}
\usepackage{graphicx}
\usepackage{dcolumn}
\usepackage{bm}
\usepackage{color}
\usepackage{units}
\usepackage{comment}
\usepackage{siunitx}
\begin{document}

\title{The ``isothermal" compressibility of active matter}

\author{Austin R. Dulaney}
 \affiliation{Division of Chemistry and Chemical Engineering, California Institute of Technology}
\author{Stewart A. Mallory}%
\affiliation{Division of Chemistry and Chemical Engineering, California Institute of Technology}
\author{John F. Brady}
\email{jfbrady@caltech.edu}
\affiliation{Division of Chemistry and Chemical Engineering, California Institute of Technology}%

\date{\today}

\begin{abstract}
We demonstrate that the mechanically-defined ``isothermal'' compressibility behaves as a thermodynamic-like response function for suspensions of active Brownian particles. 
The compressibility computed from the active pressure---a combination of the collision and unique swim pressures---is capable of predicting the critical point for motility induced phase separation, as expected from the mechanical stability criterion. 
We relate this mechanical definition to the static structure factor via an active form of the thermodynamic compressibility equation and find the two to be equivalent, as would be the case for equilibrium systems. 
This equivalence indicates that compressibility behaves like a thermodynamic response function, even when activity is large. 
Finally, we discuss the importance of the phase interface when defining an active chemical potential.
Previous definitions of the active chemical potential are shown to be accurate above the critical point but breakdown in the coexistence region. 
Inclusion of the swim pressure in the mechanical compressibility definition suggests that the interface is essential for determining phase behavior.

\end{abstract}

\maketitle

\section{\label{sec:intro}Introduction}

Response functions are central to thermodynamics and the study of critical phenomena.
These quantities, which are those most frequently probed in experiment (or simulation), include various heat capacities, compressibilities, and magnetic susceptibilities.
Each response function serves as a metric on how a specific state variable changes as other independent state variables are varied under controlled conditions. 
For example, isothermal compressibility, one of the more prominent response functions and the focus of this study, is a measure of the relative volume change of a system in response to a change in pressure at constant temperature.
Isothermal compressibility has played a central role in unraveling the confounding properties of water and more generally served as a means for identifying novel phase transitions in complex fluids.~\cite{Debenedetti2020SecondWater, Bolmatov2013, Stanley1971IntroductionPhenomena}
For an isotropic, homogeneous fluid at equilibrium the isothermal compressibility takes a simple form

\begin{equation}
\chi_{_T}=-\frac{1}{V} \left( \frac{\partial V}{\partial \Pi} \right)_T=\frac{1}{n} \left( \frac{\partial n}{\partial \Pi} \right)_T,
\label{eq:compress_def}
\end{equation}

\noindent where $V$ is the volume of the system, $\Pi$ is pressure, and $n$ is the number density defined by $n=N/V$ with $N$ being the number of particles in the system.

From a statistical mechanical perspective, response functions offer a systematic way of characterizing the magnitude of fluctuations and correlation lengths in a system.
In this context, the isothermal compressibility is a measure of local density fluctuations.
It is straight-forward to show for an isotropic, homogeneous, thermodynamic system that the isothermal compressibility is given by

\begin{equation}
nk_{B} T\, \chi_{_T}=\frac{\langle (\Delta N)^2\rangle}{\langle N \rangle}= \frac{(\langle N^2 \rangle-\langle N \rangle^2) }{\langle N \rangle},
\label{eq:compress_fluct}
\end{equation}

\noindent where $k_{B}T$ is the thermal energy scale and $\langle (\Delta N)^2\rangle$ is the variance in number density.~\cite{Reichl2016APhysics}
As this result can only be derived in the grand canonical ensemble, $N$ here is interpreted as the number of particles in a subsystem of macroscopic dimension $V$, which is in equilibrium with a much larger thermodynamic system.

\begin{figure}[t!]
\includegraphics[width=0.48\textwidth]{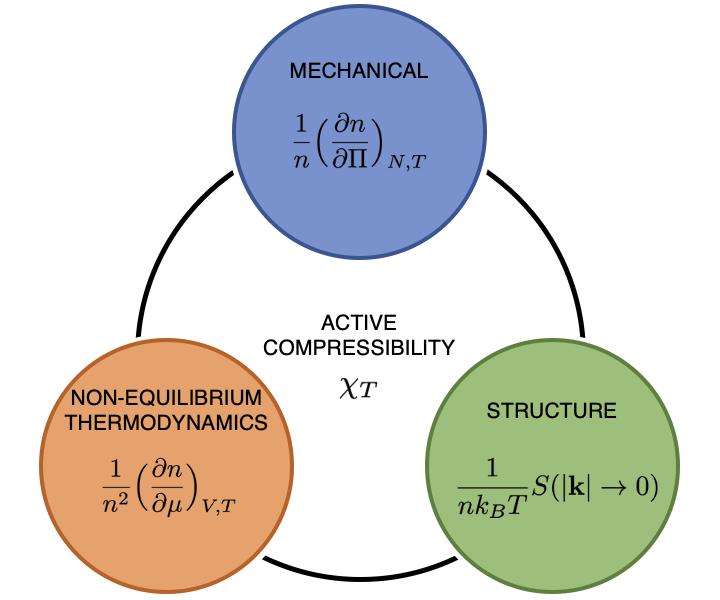}
\caption{\label{fig:compress_diagram}A postulated diagram of the compressibility in purely active systems with each method of calculation defined in the traditional thermodynamic sense.}
\end{figure}

Equivalently, one can compute isothermal compressibility directly from the system microstructure---as is often done in the study of liquids---via the compressibility equation

\begin{equation}
nk_{B} T\, \chi_{_T} = \lim_{\pmb{k} \rightarrow 0} S(\pmb{k}) = \left(1+n \int [g(r)-1] d\pmb{r} \right),
\label{eq:compress_eqn}
\end{equation}

\noindent where $g(r)$ is the radial-distribution function and S(\pmb{k}) is the static structure factor. 
Finally, one can also compute the compressibility by means of the free energy, or more precisely, from the chemical potential $\mu$:

\begin{equation}
    \chi_{T} = \frac{1}{n^2} \left( \frac{\partial n}{\partial \mu} \right)_{V,T},
    \label{eq:chem_pot_compress}
\end{equation}

\noindent where the partial derivative is taken at constant volume and temperature. 
For thermodynamic consistency to hold in equilibrium systems, we have equivalency of these three methods: mechanical, thermodynamic, and structural (see Fig.~\ref{fig:compress_diagram}).

In this study, we explore whether the same equivalence and consistency exist for an important class of nonequilibrium systems---active Brownian particles (ABPs). 
The motivation for carrying out this work is multifold. 
ABPs have become a popular minimal model for understanding the behavior of active or self-propelled colloids, bacteria, and other living systems. 
The defining characteristic of an active colloid, which makes it unique relative to its passive Brownian counterpart, is the driven and persistent nature of its motion. 
This difference in dynamics is responsible for a wealth of interesting and novel behaviors including spontaneous clustering,~\cite{Palacci2013LivingSurfers} swarming, and motility-induced phase separation.~\cite{Zottl2016EmergentColloids,Takatori2015a, Fily2012AthermalAlignment, Bechinger2016, Klamser2018ThermodynamicMatter}
For this reason, suspensions of active colloids have garnered interest from the material science and engineering community as they represent a potentially innovative approach to directed transport, self-assembly, and material design at the microscale.~\cite{Bechinger2016, Mallory2018}

The collective behavior of these active matter systems is incredibly rich and has aided the development of new nonequilibrium theories.~\cite{Berthier2017HowTransitions, Paliwal2017Non-equilibriumParticles, Nandi2017NonequilibriumParticles, Feng2017ModeParticles}
Motility-induced phase separation (MIPS) has been a particular focal point for many in the active matter community.
Surprisingly, a suspension of active colloids interacting solely through their excluded volume undergoes a nonequilibrium phase transition into a dilute and dense phase–akin to liquid-vapor coexistence in a typical equilibrium liquid.~\cite{Fily2012AthermalAlignment, Takatori2015a, Prymidis2016Vapour-liquidFluid}
A great deal of effort has gone towards developing theories capable of deducing the coexistence criterion for MIPS.
Significantly less attention has been paid to the behavior of active colloids above the critical point, where there remain a number of open and important questions.~\cite{Paoluzzi2016CriticalMatter, Partridge2018CriticalClass, Siebert2018CriticalParticles}

Due to advances in experimental methods, there has been a resurgence of interest in properties of equilibrium systems in the supercritical phase.
The highly tunable behavior of molecular supercritical fluids has lead to a number of industrial applications and the introduction of new theoretical concepts in liquid state theory such as the Fisher–Widom line, the Widom line, and the Frenkel line.~\cite{McMillan2010, Herrero2010SupercriticalApplications, Bolmatov2013}
These lines, which are identified using various thermodynamic response functions (for instance, the Widom line is often identified by a peak in the isobaric heat capacity), delineate characteristic regions within the supercritical region of the phase diagram.
It remains to be seen whether these ideas can be extended to suspensions of active colloids. 
Given the similarity of the MIPS transition with liquid-gas condensation, there is some optimism that supercritical active fluids can play as versatile a role as their molecular counterparts.

A first step toward this aim is to consolidate the notion of isothermal compressibility for active suspensions.
Like many other microscopic systems, the structure factor has become an important diagnostic for active systems. 
In one of the earliest papers regarding MIPS, Fily et al. made use of the static structure factor to characterize the phase behavior of ABPs about the MIPS critical point.~\cite{Fily2012AthermalAlignment}
A recent study by Chakraborti et al. introduced a notion of compressibility for active systems.~\cite{Chakraborti2016AdditivityParticles}
By making use of a large deviation framework and assuming the property of additivity, they define a nonequilibrium chemical potential $\mu$ as the change in a nonequilibrium free energy required to insert a particle into the system---as is done in equilibrium thermodynamics. Using this chemical potential and the partition function for a given subvolume of the system, they derived the following expression for the compressibility:

\begin{equation}
    \frac{\partial n}{\partial \mu} = \lim_{\mathcal{V} \rightarrow \infty}
    \frac{1}{\mathcal{V}}(\langle \Delta \mathcal{N} \rangle^2)=\lim_{\mathcal{V} \rightarrow \infty} \frac{1}{\mathcal{V}}(\langle \mathcal{N}^2 \rangle-\langle \mathcal{N} \rangle^2),
    \label{eq:chakraborti_compress}
\end{equation}

\noindent where $n$ is the system density and $\mathcal{N}$ is the number of particles within a subsystem of volume $\mathcal{V}$.
This formulation is similar to the one used in the grand canonical ensemble (see Eq. ~\eqref{eq:compress_fluct}).
Through careful consideration of the number fluctuations $\langle \Delta \mathcal{N} \rangle^2$ within a given subvolume, the authors were able to predict the onset of MIPS by looking at the fluctuations as a function of system density for increasing levels of activity.~\cite{Chakraborti2016AdditivityParticles}
This study raises the natural question as to whether the compressibility can be computed through the other aforementioned methods for thermodynamic systems and whether the connections between these definitions---as depicted in Fig.~\ref{fig:compress_diagram}---exist for active systems.

Using a combination of large-scale simulation and analytical theory, we focus on characterizing the compressibility of a suspension of active Brownian disks.
The manuscript is organized as follows. 
In section \ref{sec:methods}, we define our implementation of the active Brownian particle model and discuss all relevant details to performing large-scale simulations.
In section \ref{sec:mechanical_compress}, we introduce the notion of pressure and ``isothermal'' compressibility in active systems and directly compute these quantities from large scale simulation data for a wide range of volume fractions and activities. 
We explicitly demonstrate the divergence of the isothermal compressibility as the MIPS critical point is approached. 
In section \ref{sec:sq}, we motivate the validity of calculating isothermal compressibility from the definition of the structure factor for active Brownian particles, which provides an independent approach to computing compressibility directly from the fluid structure. 
In section \ref{sec:compare}, we present a comparison between the structural and mechanical definitions for compressibility and address the relation to an active chemical potential.
Lastly, we summarize our work and discuss future directions in section \ref{sec:conclusion}.

\section{\label{sec:methods} Simulation Methods}

We consider a suspension of monodisperse, athermal active particles of radii $a$.
The active motion is characterized by an intrinsic swim velocity $U_{0}\mathbf{q}$---where $\mathbf{q}$ is the particle orientation---and a timescale for reorientation $\tau_{R}$.
We evolve the system forward in time using overdamped Langevin dynamics 

\begin{align}
    0 = -\zeta \mathbf{U}_{i} + \mathbf{F}^{swim}_{i} + \sum_{i \neq j} \mathbf{F}^{P}_{ij},
    \label{eq:langevin_t}
    \\
    0 = -\zeta_{R} \mathbf{\Omega}_{i} + \mathbf{L}^{R}_{i},
    \label{eq:langevin_r}
\end{align}

\noindent where $\mathbf{F}^{P}_{ij}$ is the interparticle force for pair $ij$, ${\mathbf{F}^{swim}=\zeta U_{0}\mathbf{q}}$ is the swim force, $\mathbf{\Omega}_{i}$ is the angular velocity of particle $i$, and $\zeta$ and $\zeta_{R}$ are the translational and rotational drags, respectively.
The angular velocity relates to the evolution of particle orientations via ${\partial \mathbf{q}_{i} / \partial t = \mathbf{\Omega}_{i} \times \mathbf{q}_{i}}$.
Normalizing position and time in Eqs.~\eqref{eq:langevin_t} and \eqref{eq:langevin_r} by $a$ and $\tau_{R}$, respectively, gives rise to the nondimensional reorientation P\'eclet number ${Pe_{R} \equiv a/(U_{0}\tau_{R})}$, which measures the ratio of a particle's size to its run length $l=U_{0}\tau_{R}$, the distance traveled between reorientation events.~\cite{Takatori2015a}
Here we assume reorientations occur through a stochastic torque $\mathbf{L}^{R}$ governed by white noise statistics with zero mean and variance ${2\zeta_{R}^{2}\delta(t)/\tau_{R}}$ and our particles interact via a Weeks-Chandler-Anderson (WCA) potential with cutoff radius ${r_{cut}=(2a)2^{1/6}}$.
The depth of the potential is set such that $\epsilon = 200 F^{swim}a$. 

The primary aim in this work is to understand the behavior of active systems in the supercritical region---above the critical point---and as such we explore phase space by varying $Pe_{R}$. 
To avoid introducing an additional force scale, we hold $U_{0}$ fixed and tune the persistence by varying $\tau_{R}$. 
Each simulation, unless otherwise specified, is run for 10,000$\tau_{R}$ with a total number of 40,000 particles.
All simulations were conducted using the HOOMD-Blue software package.~\cite{Anderson2008, Glaser2015}

\section{\label{sec:mechanical_compress} Mechanical Compressibility}

The mechanical pressure exerted by a suspension of active particles on its surroundings $\Pi^{act}$ can be easily computed directly from the virial as 

\begin{equation}
\begin{split}
\Pi^{act} & = \frac{1}{V}\sum_i^N\langle \bm{x}_i \cdot \bm{F}^{tot}_i \rangle,
\label{eq:virial_force}
\end{split}
\end{equation}

\noindent where $\bm{F}^{tot}_i$ is the total force acting on particle $i$ and $\bm{x}_i$ is the position of particle $i$.
The total force acting on a given particle $\bm{F}^{tot}_{i} = \bm{F}^{swim}_{i} + \bm{F}^{col}_{i}$ arises from the particle's swimming motion and interparticle collisions, respectively. 
It follows naturally that the virial can be decomposed into the individual pressure contributions

\begin{equation}
\begin{split}
\Pi^{act} & = \frac{1}{V}\sum_i^N \left[ \langle \bm{x}_i \cdot \bm{F}^{swim}_i \rangle+\langle \bm{x}_i \cdot \bm{F}^{col}_i \rangle \right] \\
        & = \frac{\tau_{R}}{2A}\sum_{i}^{N} tr(\mathbf{F}_{i}^{swim}\mathbf{U}_{i}) + \frac{1}{V}\sum_i^N \left[ \langle \bm{x}_i \cdot \bm{F}^{col}_i \rangle \right] \\
        & = \Pi^{swim}+\Pi^{coll}.
\label{eq:pressure_virial}
\end{split}
\end{equation}

\noindent The first term is the so-called swim pressure~\cite{Takatori2014} (defined via the impulse formula~\cite{Fily2018MechanicalMatter, Patch2018Curvature-dependentPhases} in the second line) and the second is the typical collisional pressure as computed from the microscopic virial.
The total pressure $\Pi^{act}$ defined this way is a state function for spherical ABPs and its definition can be extended to unconfined systems where $\Pi^{act}$ is equal to the internal bulk pressure.
From this we are able to use this pressure from simulation at different points on the $Pe_{R}-\phi$ phase diagram to calculate the mechanical compressibility.
Here, we define $\phi=v_{p}N/V$ as the particle area fraction with particle volume $v_{p}=\pi a^{2}$ since the system is in two spatial dimensions.

\begin{figure}[b]
\includegraphics[width=0.48\textwidth]{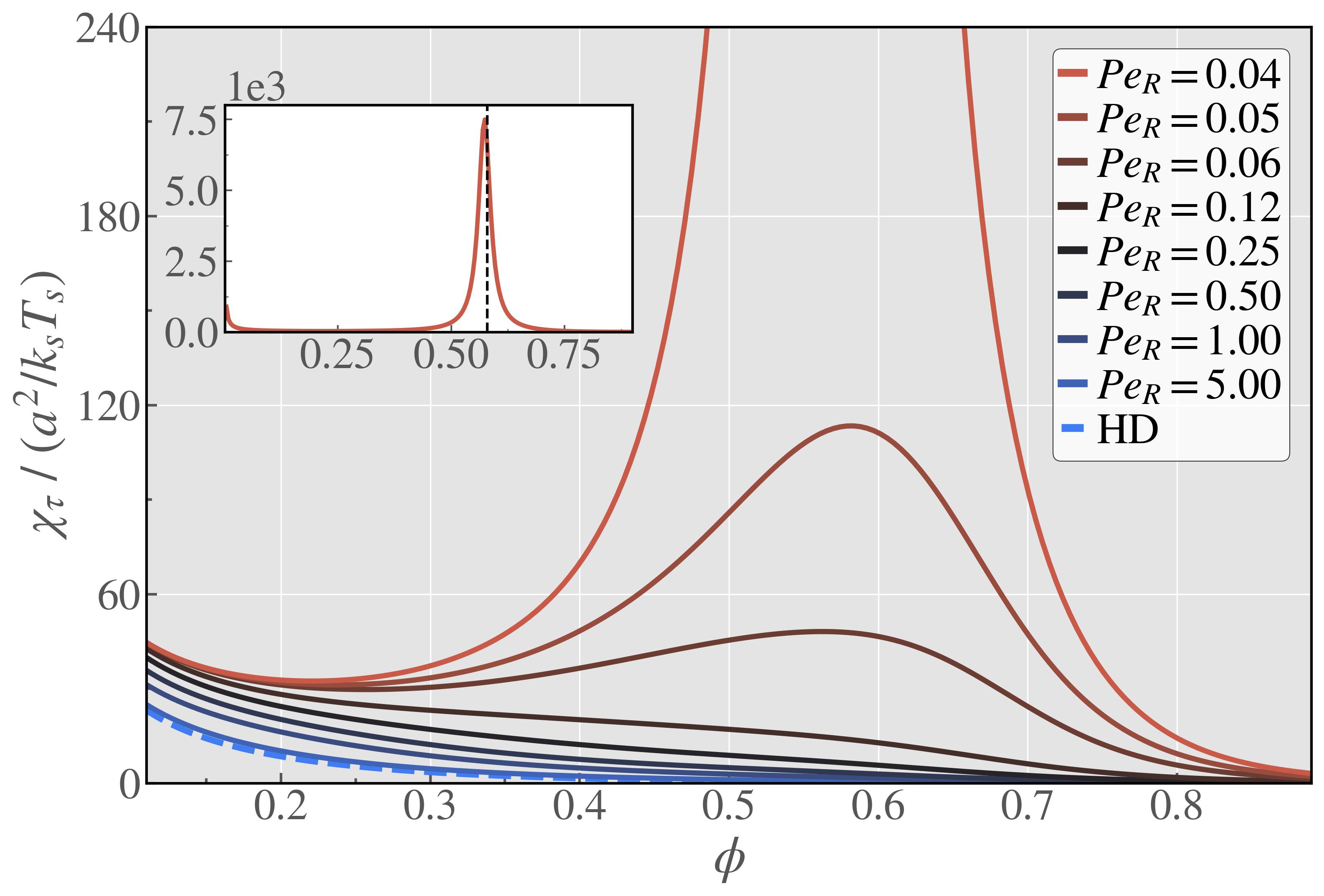}
\caption{\label{fig:k_vs_phi} Mechanical compressibility  $\chi_{\tau}$ of 2D ABPs for various $Pe_R$ as a function of volume fraction $\phi$. The compressiblity for hard disks (dashed line)---as calculated by the 10-term virial expansion---is shown for comparison. The inset shows $\chi_{\tau}$ for $Pe_{R}=0.04 \sim Pe_{R}^{crit}$.}
\end{figure}

Figure~\ref{fig:k_vs_phi} illustrates the behavior of $\chi_{\tau}$ (Eq.~\eqref{eq:compress_def}) as a function of $\phi$ for different $Pe_{R}$ values.
The compressibility is nondimensionalized by the ideal swim pressure written in terms of the activity ${k_sT_s \equiv \zeta U_{0}^2 \tau_{R}/2}$.~\cite{Takatori2014}
The hard disk compressibility (dashed line) as calculated by the first 10-terms of the virial expansion is shown for comparison.
When $Pe_{R} > 1$, the compressibility is similar to that of passive hard disk systems, even though no thermal translational motion is present in the system.
When $Pe_{R} < 1$ the compressibility becomes nonmonotonic and eventually diverges.
The $Pe_{R}$ value that is close to divergence is shown in the inset of Fig. \ref{fig:k_vs_phi}. 
Here we note that the maximum compressibility changes by several orders of magnitude for a fractional change in $Pe_{R}$ and thus we take this value to be the point of divergence.
The divergence of $\chi_{\tau}$ corresponds with the critical point, as supported by inspection of our simulations and the critical point presented by Takatori and Brady,~\cite{Takatori2015a} with a critical P\'eclet number ${Pe_{R}^{crit} \sim 0.04}$ and volume fraction ${\phi^{crit} \sim 0.58}$.

\section{\label{sec:sq}  Structure Factor}

For equilibrium systems, the low wavenumber limit of the structure factor is related to isothermal compressibility via the compressibility equation (Eq.~\eqref{eq:compress_eqn}). But this form of the compressibility equation is ill-defined in a purely active system as there is no notion of thermal energy. 
However, we can define the active compressibility via

\begin{equation}
    S(|\mathbf{k}|\rightarrow0) = n k_{s}T_{s} \chi_{\tau},
    \label{eq:new_compress_eq}
\end{equation}

\noindent where the thermal energy $k_B T$ has been replaced with the activity $k_sT_s$ as this is the relevant energy scale in the system.
The form of Eq.~\eqref{eq:new_compress_eq} comes from a mechanical argument relating the static structure factor to the particle flux, following a similar framework outlined in~\citet{Leshansky2005DynamicSuspensions}.
(The full derivation can be found in Appendix \ref{sec:app_act_compress}.) Equation~\eqref{eq:new_compress_eq} can equivalently be written in terms of the radial distribution function $g(r)$

\begin{equation}
    n k_{s}T_{s} \chi_{\tau} = 1 + n \int_{V}[g(r)-1] d\mathbf{r},
    \label{eq:rad_dist_func}
\end{equation}

\noindent where the integral is over the volume of the system. 
The definition of static structure factor does not rely on the detailed microscopic dynamics. 

\begin{figure}[t]
\includegraphics[width=0.48\textwidth]{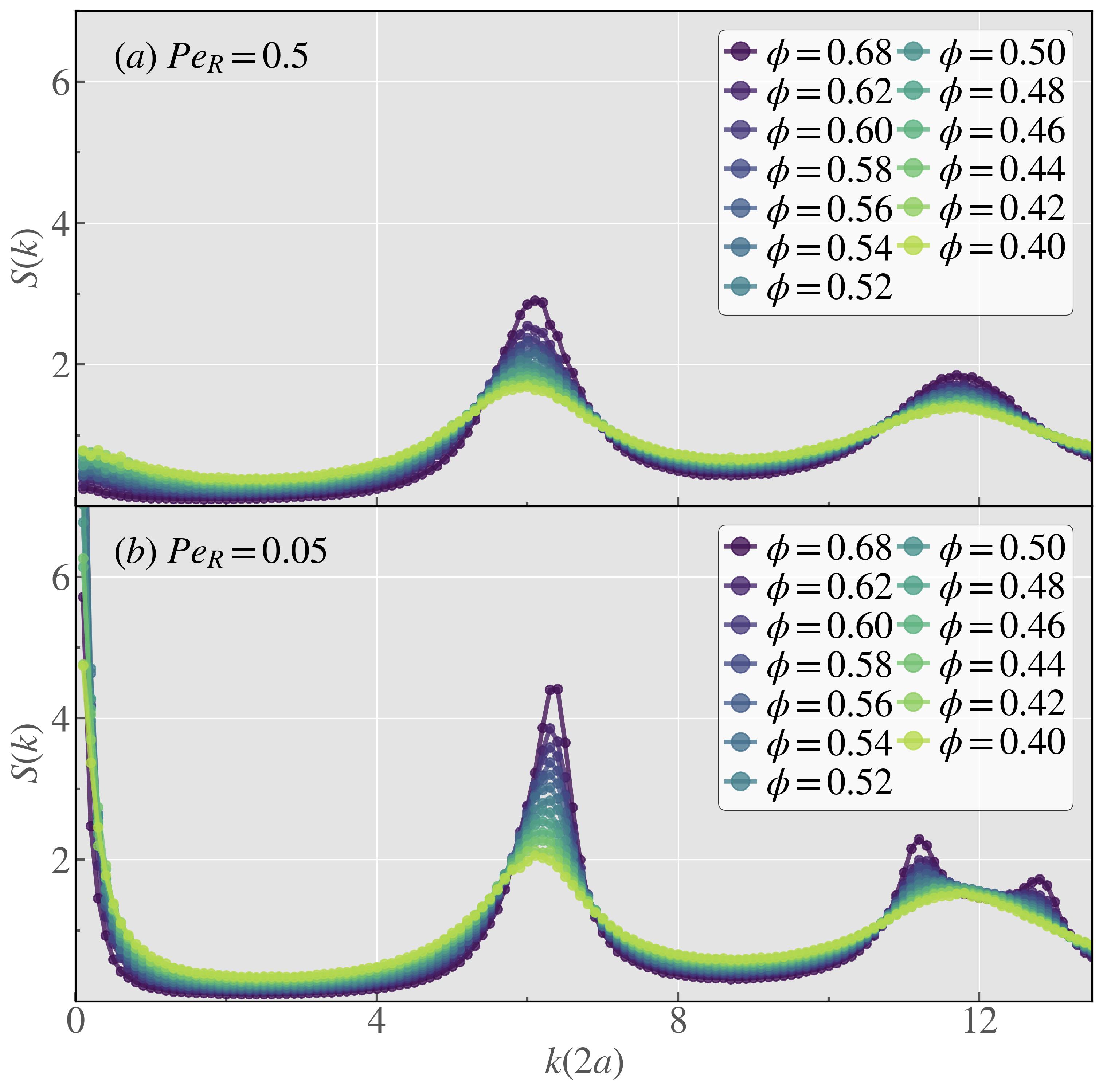}
\caption{\label{fig:struc_stack} The static structure factor for a suspension of active Brownian particles over a range of volume fractions at $(a)$ $Pe_R = 0.5$ and $(b)$ $Pe_R = 0.05$.}
\end{figure}

The static structure factor for two different values of $Pe_{R}$ over a range of volume fractions is presented in Fig.~\ref{fig:struc_stack}. 
In weakly active systems, $Pe_R \sim 1$, the structure factor behaves similarly to that of a passive system and matches well with the results presented by~\citet{DeMacedoBiniossek2018StaticDisks} (see Fig.~\ref{fig:struc_stack}(a)). 
At this level of activity $S(|\mathbf{k}|\rightarrow 0)$ never diverges.
However, as the activity increases towards the critical activity level (Fig. \ref{fig:struc_stack}(b)) $S(\mathbf{k})$ begins to diverge as $|\mathbf{k}|\rightarrow 0$ at the same critical density that was predicted in Fig. \ref{fig:k_vs_phi}, indicating a phase transition. 
This result is not surprising as the structure factor in this limit is a measure of the long-range density fluctuations and should coincide with the mechanical compressibility (Eq.~ \eqref{eq:compress_def}) as we approach the critical point. 

\section{\label{sec:compare} Compressibility Comparison}

\begin{figure}[tpb]
\includegraphics[width=0.48\textwidth]{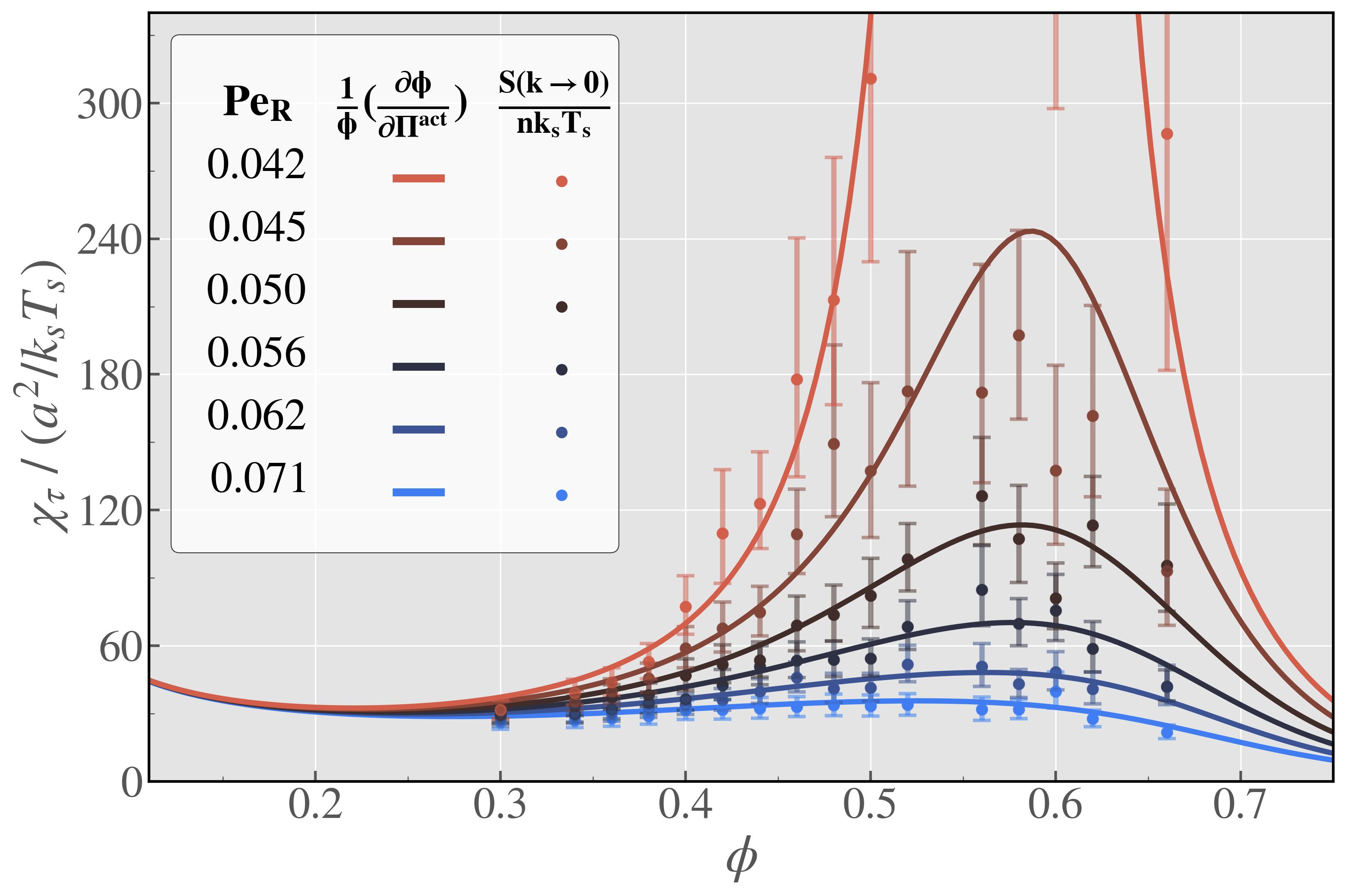}
\caption{\label{fig:compress_compare} Compressibility of active systems at varying levels of activity $Pe_{R} = 0.071,...,0.042$ computed mechanically from derivatives of the pressure (solid lines) and structurally using the compressibility equation (symbols).}
\end{figure}

Thus far we have shown that compressibility computed from the fluid structure matches the mechanical definition near the critical point, but we would like to know how the two compare throughout the homogeneous supercritical regime for active fluids.
Figure~\ref{fig:compress_compare} presents the compressibility computed via variations in the active pressure (solid lines; Eq.~\eqref{eq:compress_def}) and from the static structure factor through the use of Eq.~\eqref{eq:new_compress_eq} (symbols).
Due to the finite size of our simulations, the Nyquist sampling frequency dictates the minimum wavenumber value that can be probed $k_{min} = 2\pi / L$, where L is the size of the system. 
Therefore, the structure factor in the small wavenumber limit $\bm{k}\rightarrow 0$ was fit using the expansion $S(\bm{k})=S(0)/(1+\xi_{OZ}^{2}k^{2})$, where $S(0)$ and $\xi_{OZ}$ were used as fitting parameters, with $\xi_{OZ}$ being the Ornstein-Zernike correlation length.~\cite{Fily2012AthermalAlignment, Stopper2019OnFluids}
The error bars represent the $95\%$ confidence interval for the fitting parameters. 
The error increases as activity increases due to larger long-range density fluctuations as the critical point is approached. 
The uncertainty resulting from these fluctuations can be reduced with larger system size as this lowers the minimum possible sampling frequency.

The strong agreement between the different definitions of compressibility shows that there is meaning in the thermodynamic relations---even though the system is far from equilibrium---when using the appropriate energy scale. 
This is also evidence that compressibility behaves as a traditional thermodynamic response function even for an active system. 
The notion that compressibility gives relevant information for the phase behavior of the system also implies that the active pressure is the relevant quantity necessary to construct an equation of state for active disks.

Until now we have not considered the proposed non-equilibrium definition via the chemical potential for compressibility. 
Chakraborti \textit{et al.} have shown---through the property of additivity---the existence of a general scalar $\mu(n)$ which is tied to density fluctuations in the system.~\cite{Chakraborti2016AdditivityParticles}
As such, this scalar can be related to the compressibility through the fluctuations and consequently to the pressure, giving rise to a ``thermodynamics" of active matter (outlined in orange in Fig. \ref{fig:compress_diagram}). 
From Eqs. \eqref{eq:compress_def}, \eqref{eq:compress_fluct}, \eqref{eq:chem_pot_compress}, and \eqref{eq:chakraborti_compress} it follows that

\begin{equation}
 n \frac{\partial \mu}{\partial n} = \frac{\partial \Pi}{\partial n},   
\end{equation}

an equation first proposed by~\citet{Takatori2015a} based on arguments of particle flux driven by stress gradients (see Appendix \ref{sec:app_act_compress}).~\footnote{Takatori and Brady considered ABPs in an incompressible fluid and therefore had a factor of $(1-\phi)$ accounting for the flux of fluid in response to a flux of particles.}

This active chemical potential $\mu$ predicts the critical point~\cite{Takatori2015a} and the existence of a binodal, but it does not accurately predict the location of the binodal.
This expression for the chemical potential implies a Maxwell construction and overestimates the coexistence pressure in the two-phase region.~\cite{Paliwal2018ChemicalState} 
This is a very interesting and surprising result. 
Since the chemical potential agrees with the mechanical compressibility above the critical point,~\cite{Takatori2015a, Chakraborti2016AdditivityParticles} it implies that a thermodynamic relation can be defined and used. 
However, the same relation does not hold below the critical point.
The inaccuracy of the binodal prediction indicates that the standard Gibbs-Duhem relation is inadequate for phase coexistence and that an additional contribution is needed. 
Using active pressure as the equation of state suggests that the interface between the two phases is an essential component for determining phase behavior---due to the jump in swim pressure at the interface---and acts as an extensive property for active Brownian particles, while it does not for passive particles. 
This appears to necessitate inclusion of contributions from the phase interface when constructing an active chemical potential to arrive at the correct coexistence pressure in the two-phase region. 

\section{\label{sec:conclusion} Conclusions and Future Work}

We have computed the compressibility for an athermal, active suspension mechanically---using pressure---and structurally from the static structure factor. 
In order to compute compressibility for active systems from structure we have utilized a mechanical argument to motivate an active form of the compressibility equation. 
From this we have shown that compressibility behaves like a thermodynamic response function, as it does in equilibrium systems, so long as the swim pressure is accounted for in the total system pressure. 
As activity varies, the compressibility continuously deviates from the 2D hard disk behavior and diverges at the onset of MIPS, thus reinforcing the idea that the compressibility behaves like a response function and can be used to determine phase behavior. 

We have also discussed the existence of an active chemical potential which is linked to fluctuations in the system as shown in Chkraborti \textit{et al}.~\cite{Chakraborti2016AdditivityParticles}
This active chemical potential, while useful in the region before the onset of phase separation, does not accurately capture the location of the binodal. 
This result hints at the importance of the phase interface to determine behavior in active systems, as speculated by \citet{Solon2018GeneralizedMatter}, a surprising requirement not observed in passive systems.

While our focus has been on compressibility, its behavior as a thermodynamic response function suggests that other response functions are worth exploring in active systems. 
Compressibility serves as a natural starting point as it can be mechanically defined, but perhaps there are active analogues to other familiar thermodynamic response functions, especially considering the evidence for a non-equilibrium chemical potential and thus a non-equilibrium free energy.
There have been predictions to the form of an active heat capacity,~\cite{Takatori2015a} but to our knowledge there have been no further explorations into active analogues to thermodynamic response functions.

\section*{Data Availability}
The data that support the findings of this study are available from the corresponding author upon reasonable request.

\begin{acknowledgments}
We gratefully acknowledge the support of NVIDIA Corporation with the donation of the Titan V GPU used for this research. 
S.A.M. acknowledges support by the Arnold and Mabel Beckman Foundation. 
J.F.B acknowledges support by the National Science Foundation under Grant No. CBET-1803662. 
\end{acknowledgments}

\appendix*

\section{\label{sec:app_act_compress} Active Compressibility Equation}
Here we present the motivation for Eq. \ref{eq:new_compress_eq}, the active compressibility equation.
Following the derivation outlined by~\citet{Leshansky2005DynamicSuspensions}, we begin with the dynamic structure factor (DSF) ${F(\bm{k},t) = \langle \sum_{\alpha,\beta}\exp[i\bm{k}\cdot(\bm{x}_{\alpha}(t)-\bm{x}_{\beta}(0))]\rangle/N}$, where $\bm{k}$ is the wavenumber, the sum is over particle pairs $(\alpha, \beta)$, and the angular brackets denote an ensemble average. 
It has been shown that the collective diffusivity of a suspension can be obtained by computing the time derivative of the DSF via $\dot{F}-i\bm{k}\cdot \tilde{\bm{U}}^{*}F = -\bm{k}\cdot \bm{\hat{D}}^{coll}(\bm{k})F$, where $\tilde{\bm{U}}^{*}$ is the bulk average velocity at an arbitrary point in the suspension and $\bm{\hat{D}}^{coll}(\bm{k})$ is the Fourier transform of the collective diffusivity.~\cite{Leshansky2005DynamicSuspensions}
In the long wavelength limit (when density fluctuations persist for longer than the size of a particle), as $\bm{k} \rightarrow 0$, $F(\bm{k},t) \sim S(\bm{k})$ and $\bm{\hat{D}}^{coll}$ is given by 
\begin{equation}
\begin{aligned}
\bm{\hat{D}}^{coll}= \lim_{\bm{k}\rightarrow 0} \frac{1}{NS(\bm{k})} \langle \sum_{\alpha,\beta} \bm{M}_{\alpha \beta} e^{i\bm{k}\cdot(\bm{x}_{\alpha}(t)-\bm{x}_{\beta}(t)} \rangle,
\end{aligned}
\end{equation}

\noindent where $\bm{M}_{\alpha \beta} = \int_{t'=0}^{t} \bm{U'}_{\alpha}(t)\bm{U}_{\beta}(t')dt'$ is the mobility and $\bm{U'}_{\alpha}=\bm{U}_{\alpha} - \langle \bm{U} \rangle$ is the configuration dependent velocity fluctuation of particle $\alpha$, as presented by~\citet{Leshansky2005DynamicSuspensions}.
The velocities can be decomposed into contributions from interparticle interactions and swimming $\bm{U}_{\alpha} = \bm{U}^{P}_{\alpha} + \bm{U}^{swim}_{\alpha}$, which allows the velocity correlation function to be decomposed into interparticle-interparticle, swim-swim, and swim-interparticle components. 
The swim-swim correlation function results in the swim diffusivity $\bm{D}^{swim} = U_{0}^{2}\tau_{R}/2 \bm{I}$~\cite{Takatori2014} because only self-terms are correlated, and the interparticle-interparticle correlation goes to $0$ for pairwise interactions. This gives

\begin{equation}
\begin{aligned}
    \bm{\hat{D}}^{coll} &= \frac{1}{S(\bm{k}\rightarrow 0)}\Big(D^{swim}\bm{I}
    \\
    &+ \frac{1}{N} \langle\sum_{\alpha,\beta}\int \bm{U}_{\alpha}^{P}(t)\bm{U}_{\beta}^{swim}(t')dt'\rangle \Big).
\end{aligned}
\label{eq:coll_diff_ssf}
\end{equation}

\noindent It is important to note that were this a system of passive Brownian particles, then $D^{swim}$ would be replaced by $D_{T}$ and Brownian contributions would replace those from swimming in the above correlation functions, with the interparticle-Brownian velocity correlation function being $0$. 
For simplicity we will assume the second term on the right hand side of Eq. \ref{eq:coll_diff_ssf} is zero as it would be for Brownian motion.

Now we can relate $\bm{D}^{coll}$ to the pressure through a suspension momentum balance

\begin{equation}
    0=-\zeta n(\bm{u}_{p}-\langle\bm{u}\rangle) + \nabla \cdot \bm{\sigma}^{p},
    \label{eq:mom_balance}
\end{equation}

\noindent where $n(\bm{u}_{p}-\langle\bm{u}\rangle)=\bm{j}^{rel}$ is the relative flux, $\bm{u}_{p}$ is the particle velocity, $\langle \bm{u} \rangle = \phi \bm{u}_{p}+(1-\phi)\bm{u}_{p}$ is the suspension averaged velocity, $\bm{u_{f}}$ is the fluid velocity, and $\bm{\sigma}^{p} = -\Pi\bm{I}$ is the particle stress written in terms of pressure. 
The relative flux is representable as a generalized Fick's law $\bm{j^{rel}}=-\int\bm{D}^{coll}(\bm{x}-\bm{x'})\cdot \nabla n(\bm{x}',t)d\bm{x}'$ with a nonlocal diffusivity given by the collective diffusivity. 
Substituting this definition into Eq. \ref{eq:mom_balance} gives

\begin{equation}
\begin{aligned}
    -\int\bm{D}^{coll}(\bm{x}-\bm{x}') \cdot \nabla n(\bm{x}',t)d\bm{x}' &= \frac{1}{\zeta}\nabla \cdot \bm{\sigma}^{p}
    \\
    &= \frac{1}{\zeta}\frac{\partial \bm{\sigma}^{p}}{\partial n} \cdot \nabla n.
    \label{eq:nonlocal_coll_diff}
\end{aligned}
\end{equation}

\noindent Combining Eq. \ref{eq:coll_diff_ssf} and \ref{eq:nonlocal_coll_diff} in the limit $\bm{k}\rightarrow 0$ gives

\begin{equation}
\begin{aligned}
    \bm{\hat{D}}^{coll}(\bm{k}\rightarrow 0) = \frac{1}{\zeta} \Big(\frac{\partial \Pi}{\partial n} \Big)\bm{I} = \frac{D^{swim}\bm{I}}{S(\bm{k}\rightarrow 0)}.
\end{aligned}
\end{equation}

\noindent Using the definition for mechanical compressibility results in the well known compressibility equation~\cite{Hansen2013TheoryEdition, Stanley1971IntroductionPhenomena, Leshansky2005DynamicSuspensions}

\begin{equation}
    S(\bm{k}\rightarrow 0) = n\zeta D^{swim} \chi_{\tau} = n k_{s}T_{s} \chi_{\tau},
\end{equation}

\noindent expressed in terms of the activity $k_{s}T_{s}$ instead of the thermal energy $k_{B}T$, as presented in Section \ref{sec:sq}. A more detailed discussion regarding this derivation and its origins can be found in~\citet{Leshansky2005DynamicSuspensions}.


%

\end{document}